\documentclass[fleqn,usenatbib]{lib/mnras}

\usepackage{newtxtext,newtxmath}
\usepackage[T1]{fontenc}

\DeclareRobustCommand{\VAN}[3]{#2}
\let\VANthebibliography\thebibliography
\def\thebibliography{\DeclareRobustCommand{\VAN}[3]{##3}\VANthebibliography}

\usepackage{natbib}
\usepackage{microtype}
\usepackage{epsfig}
\usepackage{scrtime}
\usepackage{graphicx}	

\usepackage{amsmath}	
\usepackage{amssymb}	
\usepackage{float}
\usepackage[caption = false]{subfig}
\usepackage{multirow}
\usepackage{rotating}
\usepackage{ulem}
\usepackage{booktabs}
\usepackage{caption} 
\usepackage{threeparttable} 
\usepackage{threeparttablex}
\usepackage{fancyhdr} 
\usepackage{longtable}
\usepackage{xcolor}
\usepackage{multibib}

\newcommand{\ha}{\mathrm{H}\alpha}

\newcommand{\Oiii}{\mathrm{O\ III}}

\newcommand{\logd}{\log}
\newcommand{\hii}{H\,\textsc{ii}}

\newcommand{\lsigG}{$L - \sigma$}

\newcommand{\mincir}{\raise-3.truept\hbox{\rlap{\hbox{$\sim$}}\raise4.truept\hbox{$<$}\ }}

\title[Charting the expansion of the Universe to z$\sim$14]{
Charting the expansion of the Universe from z$\sim$0 to z$\sim$14 with HII galaxies}
\author[Ch\'avez et al.] {R.\, Ch\'avez$^{1, 2}$\thanks{E-mail: rchavezm@xanum.uam.mx}, 
R.\, Terlevich$^{3, 4, 5}$, 
A.\ L.\, Gonz\'alez-Mor\'an$^{3}$, 
S.\, Zamora$^{6}$, \newauthor
E.\, Terlevich$^{3, 4, 5}$, 
D.\, Fern\'andez-Arenas$^{7, 8}$, 
F.\, Bresolin$^{9}$,
M.\, Plionis$^{10, 11, 12}$,
S.\, Basilakos$^{11, 13, 14}$, 
R.\, Amorín$^{15}$, \newauthor
M.\, Llerena$^{16}$, 
F.\, D'Eugenio$^{17,18}$,
X.\, Ji$^{17,18}$,
J.\, Zavala$^{19}$, 
J.\, Rivera$^{20}$, 
N.\, Gómez-Cruz$^{3}$ and \newauthor
L.\, Corral-Bustamante$^{1}$
\\ \\
$^{1}$TecNM Cd. Cuauhtémoc, Av. Tecnológico 137, 31500, Cuauhtémoc, Chihuahua, México\\
$^{2}$Departamento de Física, Universidad Autónoma Metropolitana-Iztapalapa, Av. San Rafael Atlixco 186, A.P. 55-534, C.P. 09340, Ciudad de México, México\\
$^{3}$Instituto Nacional de Astrof\'\i sica, \'Optica y Electr\'onica,Tonantzintla, Puebla, M\'exico \\
$^{4}$Institute of Astronomy, University of Cambridge, Cambridge, CB3 0HA, UK \\
$^{5}$Facultad de Astronom\'\i a y Geof\'\i sica, Universidad de La Plata, La Plata, Argentina \\
$^{6}$Scuola Normale Superiore, Piazza dei Cavalieri 7, I-56126 Pisa, Italy \\
$^{7}$Canada–France–Hawaii Telescope, Kamuela, 96743 HI , USA \\
$^{8}$Planetarium La Enseñanza, Medellín, Antioquia, CP. 050022 Colombia\\
$^{9}$Institute for Astronomy, University of Hawaii, 2680 Woodlawn Drive, 96822 Honolulu, HI USA \\
$^{10}$Physics Dept., Aristotle Univ. of Thessaloniki, Thessaloniki 54124, Greece \\
$^{11}$National Observatory of Athens, Lofos Nymfon, 11852 Athens, Greece \\
$^{12}$CERIDES, Center of Excellence in Risk \& Decision Sciences, European University of Cyprus, Cyprus \\
$^{13}$Academy of Athens, Research Center for Astronomy and Applied Mathematics, Soranou Efesiou 4, 11527, Athens, Greece \\
$^{14}$School of Sciences, European University Cyprus, Diogenes Street, Engomi, 1516 Nicosia, Cyprus \\
$^{15}$Instituto de Astrof\'{i}sica de Andaluc\'{i}a (CSIC), Apartado 3004, 18080 Granada, Spain \\
$^{16}$INAF - Osservatorio Astronomico di Roma, Via di Frascati 33, 00078, Monte Porzio Catone, Italy\\
$^{17}$Kavli Institute for Cosmology, University of Cambridge, Madingley Road, Cambridge, CB3 0HA, UK\\
$^{18}$Cavendish Laboratory, University of Cambridge, 19 JJ Thomson Avenue, Cambridge, CB3 0HE, UK\\
$^{19}$Department of Astronomy, University of Massachusetts, LGRT-B 619E, 710 North Pleasant Street, Amherst, MA 01003-9305, United States\\
$^{20}$ Instituto de Radioastronom\'ia y Astrof\'isica, UNAM, Campus Morelia, C.P. 58089, Morelia, M\'exico\\
}

\begin{document}

\date{Compiled at \thistime\ hrs  on \today\ }

\pagerange{\pageref{firstpage}--\pageref{lastpage}} \pubyear{2026}

\maketitle

\label{firstpage}

\begin{abstract}
\noindent We present an updated Hubble diagram of giant extragalactic H\,\textsc{ii} regions and H\,\textsc{ii} galaxies spanning the redshift interval \(z\sim 0\) to \(z\sim 14\), extending the use of the \(L(\mathrm{H}\beta)-\sigma\) relation as an independent cosmological probe into the epoch of cosmic dawn. Our sample comprises 243 objects, including local anchor systems with primary distance measurements, previously published low- and intermediate-redshift H\,\textsc{ii} galaxies, recent JWST/NIRSpec observations, and a new subsample of extremely high-redshift H\,\textsc{ii} galaxies observed with JWST and, for the highest-redshift cases, with ALMA+JWST. We homogenise the measurements across the full sample, applying consistent corrections for instrumental and thermal broadening, extinction, and, where required, the transformation from [O\,III]-based to Balmer-line velocity dispersions. We model lensing effects of high redshift using a log-normal magnification formalism and infer cosmological parameters via a nested-sampling analysis, considering both spatially flat and non-flat versions of the \(\Lambda\mathrm{CDM}\), \(w\mathrm{CDM}\), and \(\mathrm{CPL}\) models. The \(L-\sigma\) relation remains consistent over the full redshift range, showing no evidence for significant evolution even at the highest redshifts currently accessible. For the joint anchor+H\,\textsc{ii} galaxies sample, considering the spatially flat models, we obtain \(h=0.725\pm0.040\) and \(\Omega_m=0.308^{+0.043}_{-0.053}\) for \(\Lambda\mathrm{CDM}\). Allowing a constant dark-energy equation of state yields \(w_0=-0.96^{+0.53}_{-0.21}\), while a CPL parametrisation gives \(w_0=-0.92^{+0.57}_{-0.34}\) and \(w_a=-0.48^{+0.60}_{-1.50}\), all consistent, within the uncertainties, with concordance cosmology. 
These results demonstrate that H\,\textsc{ii} galaxies offer a viable, fully independent tracer of the expansion history across almost the entire age of the Universe, opening a new avenue for testing \(\Lambda\)CDM and dark-energy evolution well beyond reionisation. 
\end{abstract}


\section{Introduction}\label{Intro}
The first direct evidence for the accelerated expansion of the Universe emerged almost three decades ago from Type~Ia supernovae (SNIa) \citep{Riess1998, Perlmutter1999}. Subsequent, concordant constraints from cosmic microwave background (CMB) anisotropies \citep[e.g.][]{Jaffe2001, Pryke2002, Spergel2007, PlanckCollaboration2014, PlanckCollaboration2016a, PlanckCollaboration2020} and baryon acoustic oscillations (BAO) \citep[e.g.][]{Eisenstein2005, Blake2011, Alam2021}, together with independent determinations of the expansion rate \citep[e.g.][]{Chavez2012, Freedman2012, Riess2016, Riess2018, Fernandez2018}, have established a dark-energy (DE) component as a fundamental element of the standard cosmological model.

A key question is whether the DE equation-of-state parameter, \(w \equiv p/(\rho c^2)\), where $c$ is the speed of light, $p$ is the pressure, and $\rho$ is the mass density of the DE component, deviates from a cosmological constant and/or varies over cosmic time \citep{PeeblesRatra1988, Wetterich1988}. Tighter constraints on \(w(z)\) and cross-validation with independent probes are crucial to establishing a robust cosmological framework that reduces systematic biases and tests extensions to \(\Lambda\)CDM \citep[e.g.][]{PlanckCollaboration2020, Alam2021, Moresco2022}.

In this context, extending the set of independent distance indicators is essential, both to increase the redshift leverage of the Hubble diagram and to test for probe-dependent systematic effects. Among the proposed complementary tracers are \hii\ galaxies (HIIGs) and giant extragalactic \hii\ regions (GEHRs). Their strong recombination lines make them observable over a wide redshift range and their physical properties allow their luminosities to be standardized.

HIIGs and GEHRs are intense, compact bursts of star formation, predominantly found in dwarf irregular hosts or in the outer disks of late-type galaxies. They are characterised by strong, narrow nebular lines powered by massive young stellar clusters or groups of clusters (super star clusters; SSCs). Our primary selection criterion is that their Balmer emission lines have the largest possible equivalent widths ( $EW({\rm H}\beta) > 50$\,\AA\ or $EW({\rm H}\alpha) > 300$\,\AA\ ). This selection ensures that the dominant stellar population is $\lesssim$5\ Myr old, thereby guaranteeing a sample free of long-term evolutionary effects \citep{Searle1972, Bergeron1977, Terlevich1981, Kunth2000, Chavez2014}. These  features, visible even at extremely high redshifts, make HIIGs/GEHRs important tracers of very recent star formation throughout cosmic history.

A strong correlation between the  Balmer-line luminosity (e.g., \(L({\rm H}\beta)\)) and the ionised gas non-thermal velocity dispersion \(\sigma\), known as the \(L\!-\!\sigma\) relation, has been established for HIIGs/GEHRs \citep{Terlevich1981, Melnick_Terlevich_Moles1988, Bordalo_Telles2011, Chavez2014}. The \(L\!-\!\sigma\) relation serves as a standardisable candle for cosmological distances  calibrated using nearby anchor systems with independently determined distances from Cepheids or the Tip of the Red Giant Branch (TRGB) \citep{Plionis2011, Chavez2012,   Chavez2016, Fernandez2018, GonzalezMoran2019, 2021MNRAS.505.1441G, Chavez2025}. 

The James Webb Space Telescope (JWST) offers a unique opportunity to extend this technique into the reionisation and cosmic-dawn eras. Here, we use the epoch of reionisation to denote approximately the interval \(6\lesssim z\lesssim10\), with CMB measurements placing its model-dependent midpoint at \(z_{\rm re}\simeq7.7\) \citep{PlanckCollaboration2020}. NIRSpec \citep{Dorner2016, Jakobsen2022, NIRSpecDocs} provides multiplexed spectroscopy in the range 0.6–5.3\,$\mu$m, enabling access to ${\rm H}\alpha$ up to \(z\!\sim\!7\) and ${\rm H}\beta$+[{\rm O\,III}]  up to \(z\!\sim\!9\)–10 \citep[e.g.][]{2024A&A...690A.288B}. Crucially, MIRI \citep{Rieke2015} spectroscopy extends coverage to 5–28 \,$\mu$m, allowing direct detections of rest-frame optical lines at \(z\!\gtrsim\!10\), including ${\rm H}\alpha$ and [{\rm O\,III}] at \(z\!\simeq\!10\)–12 \citep{Hsiao2024, Zavala2025}. JWST has already delivered spectroscopic confirmation of galaxies at \(z\!>\!12\) \citep{CurtisLake2023, Wang2023, Carniani2024, Naidu2026, Castellano2024}, demonstrating that compact, extreme line-emitting starbursts, the analogues of HIIGs, exist deep into the first few hundred Myr of cosmic evolution. 

The JWST detection of numerous extremely high-redshift HIIGs significantly advances the investigation of an epoch when the Universe was mostly neutral, matter dominated and its expansion was slowing down.

We have recently \citep{Chavez2025} extended the HIIG/GEHR Hubble diagram to \(z\!\sim\!7.5\) using JWST/NIRSpec data, demonstrating that the \(L\!-\!\sigma\) relation remains valid at extremely early epochs and, when combined with local anchors, yields independent, competitive constraints on the Hubble constant, \(H_0\), as well as $w$, consistent with contemporary SNIa and BAO inferences. Building on that foundation, JWST now unlocks rest-frame optical lines at even higher redshifts, providing the leverage to probe the expansion history during cosmic dawn.

In this paper, we extend the approach to an unprecedented redshift range, constructing an HIIG/GEHR Hubble diagram from \(z\!\approx\!0\) to \(z\!\approx\!14\). By combining JWST and ALMA spectroscopy to measure Balmer and oxygen lines across this range, we derive distance moduli that trace the Hubble flow from the local Universe to well beyond reionisation. We use these data to constrain \(\Lambda\)CDM and simple extensions with evolving \(w(z)\),
providing an independent test of the expansion history of the Universe over approximately 98\% of cosmic time.

\section{Data sets}

We assembled a working sample from the literature,  as follows:
\begin{itemize}
    \item \textit{Local anchors:} 36 nearby extragalactic objects with independently measured distance moduli from \citet{Fernandez2018} and references therein.
    \item \textit{Low redshift ($0.01 \le z \le 0.15$):}  107 HIIGs from \citet{Chavez2014}.
    \item \textit{Intermediate redshift ($0.6 \le z \le 4$):}  24 HIIGs from \citet{Erb2006}, \citet{Masters2014}, and \citet{Maseda2014};  6 VLT/X-shooter targets from \citet{Terlevich2015}; 15 Keck/MOSFIRE objects from \citet{GonzalezMoran2019} and 29 VLT/KMOS objects from \citet{2021MNRAS.505.1441G}.
    \item \textit{High redshift ($ z > 4 $):} 5 HIIGs observed with JWST/NIRSpec \citep{2024A&A...684A..87D, Eisenstein2026, Rieke2023, 2024A&A...690A.288B} and 9 HIIGs drawn from VUDS/VANDELS \citep{2023A&A...676A..53L}, for which the homogenised compilation and measurements were presented in \citet{Chavez2025}.
\end{itemize}

In this work, we also introduce a new JWST subsample that extends the Hubble diagram to redshifts greater than 14. In Table ~\ref{tab:data} we summarize the  measurements for our new subsample. We divide it into two groups: the first comprises objects observed entirely with JWST/NIRspec, while the second comprises galaxies for  which the velocity dispersions are measured from [O\,III]88 $\mu m$ with ALMA and the fluxes from rest-frame optical lines with JWST/MIRI \citep{Carniani2024,Calabro2024,Zavala2025,Schouws2025,Carniani2025, Helton2025}.

\begin{figure}
\centering\includegraphics[width=\columnwidth]
{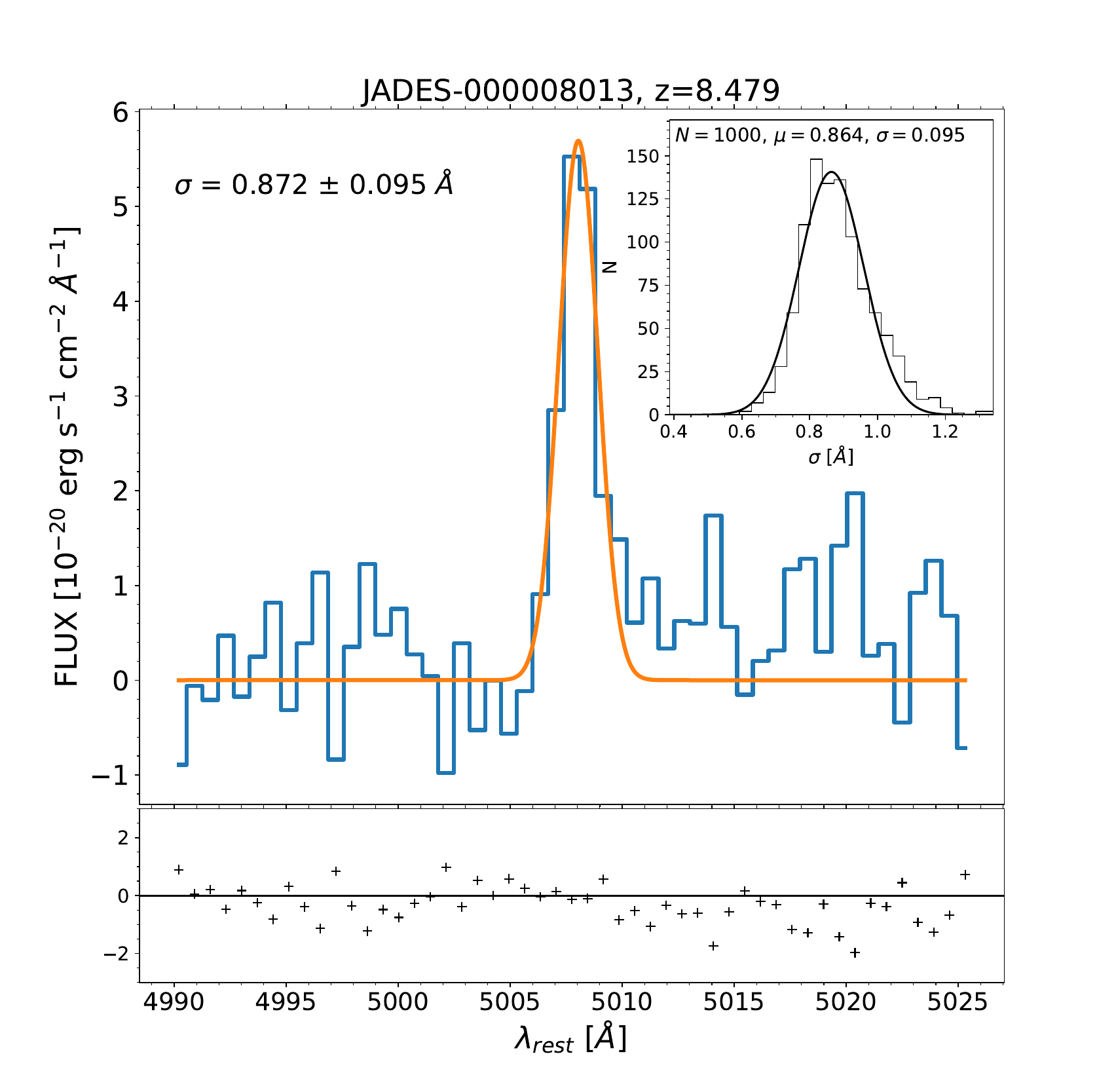}
\caption{Observed [O III]$ \lambda$5007\AA\ line profile for the target JADES-000008013. The blue line is the spectrum, the orange line is the Gaussian fit to the emission line, and the box underneath shows the residuals. The inset at the upper right corner is the Monte Carlo analysis performed on the line, with the standard deviation of the resulting distribution taken as the uncertainty of the measured FWHM of the Gaussian fit.}
\label{fig:OIII_profile_8013}
\end{figure}

Regarding the JWST/NIRSpec data, we have measured the velocity dispersion of the ionised gas from Gaussian fits to the emission lines detected in spectra observed with the  F290LP filter and the high resolution, R $\sim$ 2,700, grating G395H. In Figure \ref{fig:OIII_profile_8013}, we show the observed \mbox{[O III]}$\lambda$5007\AA\  profile for the highest redshift (z $\sim$ 8.47) target of the JWST/NIRSpec sample, JADES-000008013. At this redshift, the $\ha$ line is no longer detected in the  F290LP/G395H configuration. As in previous work and for consistency, when only [O\,III]$\lambda$5007\AA\  based velocity dispersions are available, we apply a $2.1~\mathrm{km\,s^{-1}}$ correction to transform them to the Balmer line velocity dispersion scale \citep{Chavez2016}.

We calculate the 1D non-thermal velocity dispersion $\sigma$ fitting a Gaussian profile to each emission line to determine $\sigma_{obs}$. We then subtract in quadrature from $\sigma_{obs}$ the instrumental resolution, as described in \citet{2024A&A...684A..87D}, along with the thermal and fine-structure broadening following \cite{GonzalezMoran2019} procedure.
 
 The observed H$\beta$ fluxes are corrected for aperture effects and dereddened using \citet{Gordon2003} SMC Bar attenuation law. Where available, the Balmer decrement (H$\alpha$/H$\beta$) is used to infer the extinction, and when H$\alpha$ is unavailable, we adopt the mean extinction from our previous compilation. Uncertainties are propagated throughout the entire procedure.

\begin{table*}
        \caption{Data set measurements.}
\label{tab:data}
\tabcolsep 5pt
\begin{tabular}{@{}lcccc@{}}
\toprule
Object & $z$ & $\logd \sigma$ & $\logd f (\mathrm{H}\beta)$ & $EW(\mathrm{H}\alpha)$ \\
&&($\mathrm{km\ s^{-1}}$)&($\mathrm{erg\ s^{-1}\ cm^{-2}}$)&(\r{A}, Rest-Frame)\\
\midrule
\multicolumn{5}{c}{\textbf{JWST data}} \\
JADES-10013609 & $6.92889 \pm 0.00069^{a}$ & $1.682 \pm 0.060^{d}$& $-17.83 \pm 0.19^{h}$ & ~$1847 \pm 131^{j}$ \\
JADES-00018976 & $6.32705 \pm 0.00063^{a}$ & $1.630 \pm 0.052^{d}$& $-18.43 \pm 0.18^{h}$ & ~$517 \pm 47^{j}$ \\
JADES-10005113 & $5.82084 \pm 0.00058^{a}$ & $1.503 \pm 0.049^{e}$& $-18.68 \pm 0.08^{h}$ & $1368 \pm 87^{j}$ \\
JADES-10013704 & $5.91983 \pm 0.00059^{a}$ & $1.819 \pm 0.014^{d}$& $-17.13 \pm 0.08^{h}$ & ~~$761 \pm 42^{j}$\\
JADES-00004404 & $5.76300 \pm 0.00058^{a}$ & $1.742 \pm 0.021^{d}$& $-17.80 \pm 0.20^{h}$ & $1200 \pm 66^{j}$ \\
JADES-10013905& $7.19700\pm 0.00072^{a}$& $1.724 \pm 0.042^{d}$& $-18.13 \pm 0.05^{h}$ & $ ~~~~477 \pm 156^{j}$ \\
JADES-000019342 & $5.98100 \pm 0.00060^{a}$ & $1.691 \pm 0.049^{d}$& $-18.36 \pm 0.10^{h}$ & ~~$490 \pm 41^{j}$ \\
JADES-000004297 & $6.71800 \pm 0.00067^{a}$ & $1.662 \pm 0.036^{d}$& $-18.26 \pm 0.19^{h}$ & ~~$1533 \pm 102^{j}$ \\
JADES-000008013 & $8.47900 \pm 0.00085^{a}$ & $1.686 \pm 0.065^{d}$& $-18.74 \pm 0.05^{h}$ & ~~~~$300 \pm 123^{j}$ \\
JADES-000016625 & $6.63100 \pm 0.00066^{a}$ & $1.668 \pm 0.033^{e}$& $-18.13 \pm 0.06^{h}$ & ~$2315 \pm 98^{j}$ \\
\midrule
\multicolumn{5}{c}{\textbf{ALMA+JWST data}} \\
GHZ2/GLASS-z12& $12.342 \pm 0.009^{b}$& $1.923 \pm 0.133^{f}$ & $-18.06 \pm 0.53^{i}$ & $1450 \pm 312^{k}$\\
JADES-GS-z14-0&$14.1793\pm 0.0007^{c}$& $1.668 \pm 0.116^{g}$ & $-18.36 \pm 0.72^{i}$ & $80 \pm 50^{l}$ \\
\hline
\multicolumn{5}{l}{}\\
\multicolumn{5}{l}{
$^{a}$Measured directly from JWST/NIRSpec spectra. $^b$Measured in \citet{Zavala2025} from JWST/MIRI}\\
\multicolumn{5}{l}{spectra. $^c$Measured in \citet{Schouws2025} from the [O\,III]88 $\mu m$ ALMA line detection. $^d$Measured}\\ \multicolumn{5}{l}{from the JWST/NIRSpec spectra $[\Oiii]\lambda 5007$ line and corrected by thermal and instrumental broadening.}\\ 
\multicolumn{5}{l}{$^e$Measured from the JWST/NIRSpec spectra H$\alpha$ line and corrected by thermal and instrumental broadening.}\\
\multicolumn{5}{l}{$^f$Measured by \citet{Zavala2024} from the ALMA [O\,III]88 $\mu m$ line. $^g$Measured by \citet{Schouws2025}} \\
\multicolumn{5}{l}{from the ALMA [O\,III]88 $\mu m$ line. $^h$Measured from the JWST/NIRSpec medium resolution spectra and}\\
\multicolumn{5}{l}{corrected by aperture and extinction effects. $^i$Measured from the JWST/MIRI spectra and corrected by}\\
\multicolumn{5}{l}{aperture and extinction effects. $^j$Measured from JWST/NIRSpec prism spectra.}\\
\multicolumn{5}{l}{$^k$EW(H$\beta$+[O III]) estimated by \citet{Greve2026} using the JWST/MIRI fluxes reported in \citet{Zavala2025}}\\
\multicolumn{5}{l}{and the F444W continuum photometry. $^l$Measured from JWST/MIRI spectra by \citet{Helton2025}.}
\end{tabular}
\end{table*}

\section{Constraints on cosmological parameters}
We infer cosmological parameters using the \(\mathrm{L}{-}\sigma\) distance estimator for GEHRs and HIIGs, thereby refining the framework of our earlier analyses \citep{Chavez2016, GonzalezMoran2019, 2021MNRAS.505.1441G,Chavez2025}.

The quantities measured directly for each object are the extinction-corrected emission-line flux, \(f\), and the velocity dispersion, \(\sigma\), rather than the intrinsic luminosity. The \(L-\sigma\) relation predicts the luminosity from \(\sigma\) once its slope and zero-point have been calibrated. Combining this prediction with \(L=4\pi d_L^2f\) yields an estimate of the luminosity distance and hence of the distance modulus. The local anchor objects provide the independent primary distances required to establish the absolute calibration of the relation. In the global analysis, the calibration parameters \(\alpha\) and \(\beta\) are sampled jointly with the cosmological parameters using the anchor and HIIG data. Therefore, the observed distance estimator is constructed from \(f\), \(\sigma\), and the calibrated \(L-\sigma\) relation, while the cosmological model enters through the predicted distance modulus \(\mu_\theta(z\mid\theta)\).

This section provides a single, self–contained description of likelihood, cosmological models, uncertainty propagation, bias corrections, and inference setup  to ensure reproducibility.


\textbf{Likelihood and distance estimator.}
 For each object, we compare the distance modulus predicted by a cosmological model with the distance modulus inferred from observables via  the \(\mathrm{L}{-}\sigma\) relation. The likelihood for the HIIG/GEHR sample is
\begin{equation}
    \mathcal{L}_{H} \propto \exp\!\left(-\frac{1}{2}\chi^2_{H}\right),
    \label{eq:lkh}
\end{equation}
with
\begin{equation}
    \chi^2_{H}=\sum_{n}\frac{\left[\mu_o(\log f,\log \sigma\,|\,\alpha,\beta)\;-\;\mu_{\theta}(z\,|\,\theta)\right]^2}{\epsilon^2}.
    \label{eq:chi}
\end{equation}
The observed distance modulus \(\mu_{o}\) is obtained from the \(\mathrm{L}{-}\sigma\) relation as
\begin{equation}
    \mu_o \;=\; 2.5\left(\beta\,\log \sigma + \alpha - \log f - 40.08\right),
    \label{eq:muo}
\end{equation}
where \(\alpha\) and \(\beta\) are the intercept and slope, \(\sigma\) is the line-of-sight velocity dispersion (corrected for instrumental and thermal broadening) and \(f\) is the extinction-corrected line flux. For HIIGs, \(\mu_{\theta}=5\log_{10} d_L(z,\theta)+25\) with \(d_L\) in Mpc; for the anchor objects, \(\mu_{\theta}\) is the independently measured (primary–indicator) distance modulus \citep{Chavez2016, GonzalezMoran2019, Chavez2025}.

\textbf{Cosmological models.}
We consider both spatially flat and curved cosmologies. Our baseline dark–energy equation of state (EoS) adopts the Chevallier–Polarski–Linder (CPL) form \citep{Chevallier2001, Linder2003, Peebles2003, Dicus2004, Wang2006}:
\begin{equation}
    w(z)=w_0+w_a\frac{z}{1+z},
    \label{eq:CPL_model}
\end{equation}
with \(\Lambda\)CDM recovered at \((w_0,w_a)=(-1,0)\) and \(w\)CDM at \(w_a=0\).

\emph{Flat CPL (fiducial).} For a spatially flat Universe the normalised Hubble function is
\begin{equation}
\begin{aligned}
    E^2(z,\theta) &= \Omega_{r}(1+z)^4 + \Omega_{m}(1+z)^3 \\
    &\quad + \Omega_{\mathrm{DE}}\,(1+z)^{3(1+w_0+w_a)}\exp\!\left(-\frac{3w_a z}{1+z}\right),
\end{aligned}
    \label{eq:Ez_flat}
\end{equation}
with parameter vector \(\theta=\{h,\Omega_m,w_0,w_a\}\) and \(\Omega_{\mathrm{DE}}=1-\Omega_m-\Omega_r\). We retain the radiation term \(\Omega_r\) explicitly given the redshift lever arm and we use 
\(h\equiv H_0/(100\,\mathrm{km\,s^{-1}\,Mpc^{-1}})\) as the reduced Hubble constant.

\emph{Non-flat CPL.} To allow spatial curvature, we promote the parameter set to \(\theta=\{h,\Omega_m,w_0,w_a,\Omega_k\}\) and write
\begin{equation}
\begin{aligned}
    E^2(z,\theta) &= \Omega_{r}(1+z)^4 + \Omega_{m}(1+z)^3 + \Omega_{k}(1+z)^2 \\
    &\quad + \Omega_{\mathrm{DE}}\,(1+z)^{3(1+w_0+w_a)}\exp\!\left(-\frac{3w_a z}{1+z}\right),
\end{aligned}
    \label{eq:Ez_curved}
\end{equation}
where \(\Omega_{\mathrm{DE}}=1-\Omega_m-\Omega_r-\Omega_k\). The \(\Lambda\)CDM and \(w\)CDM limits are recovered by setting \((w_0,w_a)=(-1,0)\) and \(w_a=0\), respectively.

\emph{Minimal non-flat \(\Lambda\)CDM.} When focusing on a curvature test without EoS freedom, we use the three-parameter model \(\theta=\{h,\Omega_m,\Omega_{\mathrm{DE}}\}\), with
\begin{equation}
\begin{aligned}
    E^2(z) &= \Omega_{r}(1+z)^4 + \Omega_{m}(1+z)^3 + \Omega_{k}(1+z)^2 + \Omega_{\mathrm{DE}}, \\
    \Omega_{k} &= 1 - \Omega_m - \Omega_r - \Omega_{\mathrm{DE}},
\end{aligned}
    \label{eq:Ez_nflcdm}.
\end{equation}

\emph{Distances in curved geometry.} For all cases we compute the line-of-sight comoving distance
\begin{equation}
    \chi(z)=\int_{0}^{z}\frac{dz'}{E(z',\theta)},
\end{equation}
and the transverse comoving distance \(D_M\) via the curvature-dependent expression:
\begin{equation}
    D_M(z)=\frac{c}{H_0}\,
    \begin{cases}
        \frac{1}{\sqrt{\Omega_k}}\sinh\!\left(\sqrt{\Omega_k}\,\chi\right), & \Omega_k>0 \ (\text{open}),\\[4pt]
        \chi, & \Omega_k=0 \ (\text{flat}),\\[4pt]
        \frac{1}{\sqrt{|\Omega_k|}}\sin\!\left(\sqrt{|\Omega_k|}\,\chi\right), & \Omega_k<0 \ (\text{closed}),
    \end{cases}
    \label{eq:Sk}
\end{equation}
so that the luminosity distance is \(d_L(z)=(1+z)\,D_M(z)\) \citep[see, e.g.,][]{Hogg1999}. In the flat limit, \(D_M=(c/H_0)\chi\), recovering Eq.~\ref{eq:Ez_flat} combined with our standard \(d_L\) definition.

\textbf{Uncertainty model and error propagation.}
The denominator in Eq.~\ref{eq:chi} combines statistical and systematic contributions
\begin{equation}
    \epsilon^2 \;=\; \epsilon^2_{\mu_{o},\mathrm{stat}} \;+\; \epsilon^2_{\mu_{\theta},\mathrm{stat}} \;+\; \epsilon^2_{\mathrm{sys}},
    \label{eq:epsilon}
\end{equation}
with  observational propagation for \(\mu_o\)
\begin{equation}
    \epsilon^2_{\mu_{o},\mathrm{stat}}
    = 6.25\left(\epsilon_{\log f}^2 + \beta^2\,\epsilon_{\log \sigma}^2 + \epsilon_{\beta}^2\,\log^2 \sigma + \epsilon_{\alpha}^2 \right),
    \label{eq:epsilon_mu_o_stat}
\end{equation}
where \(\epsilon_{\log f}, \epsilon_{\log \sigma}, \epsilon_{\alpha}, \epsilon_{\beta}\) are the measurement uncertainties on the respective quantities. The term \(\epsilon_{\mu_{\theta},\mathrm{stat}}\) propagates redshift errors for HIIGs and primary-distance errors for anchors. The systematic term \(\epsilon_{\mathrm{sys}}\) adds the effects of starburst age, size,
abundance/excitation, extinction curve choice and line-choice.


\textbf{Bias corrections and additional systematics.}
\textit{Malmquist bias:} for flux-limited subsamples (notably the nearby calibrators), we compute and apply the Malmquist correction following \citet{Giraud1987} as implemented in \citet{Chavez2016}. We add the corresponding uncertainty in quadrature to \(\epsilon_{\mathrm{sys}}\). \textit{Weak gravitational lensing:} we account for the redshift-dependent magnification scatter and small de-magnification bias at high \(z\) using the lognormal prescription of \citet{HolzLinder2005}, as adopted for HIIG/GEHR standard-candle applications in \citet{Plionis2011}. A detailed prescription of this treatment is provided in section \ref{sec:weak_lensing}. \textit{Aggregate systematics:} The total systematic uncertainty is
obtained by combining the individual contributions in quadrature.
As shown by \citet{2021MNRAS.505.1441G}, this treatment yields
$\chi^2_\nu\!\approx\!1$ for $\Lambda$CDM fits and limits parameter
shifts to within the quoted uncertainties.

\textbf{\(h\)-independent formulation.}
To remove explicit dependence on \(H_0\), we also adopt the \(h\)-free likelihood \citep{Nesseris2005}. First we define the dimensionless luminosity distance
\begin{equation}
    \begin{split}
    D_L(z,\theta) &= (1+z)\,
    \begin{cases}
        \frac{1}{\sqrt{\Omega_k}}\sinh\!\left(\sqrt{\Omega_k}\,\chi\right), & \Omega_k>0 \ (\text{open}),\\[4pt]
        \chi, & \Omega_k=0 \ (\text{flat}),\\[4pt]
        \frac{1}{\sqrt{|\Omega_k|}}\sin\!\left(\sqrt{|\Omega_k|}\,\chi\right), & \Omega_k<0 \ (\text{closed}),
    \end{cases}, \\ d_L(z,\theta)&=\frac{c}{H_0}\,D_L(z,\theta).
    \end{split}
\end{equation}
We then fit directly for \(D_L\), which removes \(h\) from the cosmological parameter vector; implementation details follow \citet{GonzalezMoran2019, 2021MNRAS.505.1441G}.

\textbf{Inference and priors.}
We explore parameter posteriors using the \textsc{MultiNest} nested-sampling algorithm \citep{Feroz2008, Feroz2009, 2019OJAp....2E..10F} (via \textsc{PyMultiNest}). Unless otherwise stated, we adopt uniform, uninformative priors as in \citet{2021MNRAS.505.1441G, Chavez2025}: \(h\in[0.5,1.0]\), \(\Omega_m\in[0,1]\), \(w_0\in[-2,0]\), \(w_a\in[-2,2]\), and \(\alpha\in[32.5,34.5]\), \(\beta\in[4.5,5.5]\). Nuisance \((\alpha,\beta)\) and cosmological parameters are sampled jointly; we report whether results are \(h\)-free or derived from the full likelihood of Eqs.~\ref{eq:lkh}–\ref{eq:epsilon_mu_o_stat}.


\textbf{Figure of merit.}
To facilitate comparison with the literature, we quote the figure of merit \citep{Wang2008}
\begin{equation}
    FoM \;=\; \frac{1}{\sqrt{\det\, \mathrm{Cov}(\theta_0,\theta_1,\ldots)}},
\label{eqFOM}
\end{equation}
computed from the marginalised covariance of the chosen parameter subset (e.g. \(\{w_0,w_a\}\) for CPL).

\subsection{Gravitational lensing treatment}
\label{sec:weak_lensing}

The gravitational potential of structures intervening between the
 source and the receptor affect the propagation of light from high redshifts and thus also the distance modulus of high-$z$ standard candles \citep[e.g.][]{Holz1998, HolzLinder2005, Brouzakis2008}. These studies assume a Robertson–Walker cosmological background with superimposed local inhomogeneities and include both strong and weak gravitational lensing effects. The resulting flux magnification probability distribution [$P(\mu_{a})$], obtained from Monte Carlo and ray-tracing techniques, is approximately log-normal, mean magnification of unity, a mode shifted towards demagnification, and a long high-magnification tail.
Consequently, most high-redshift sources are expected to appear
slightly demagnified while a smaller fraction become strongly magnified.
Although the detailed form of the magnification distribution
depends on the underlying cosmology, the density profile and evolutionary
phase of the intervening cosmic structures, the main features of the
$P(\mu_a)$ distribution remain robust \citep[e.g.][]{Wang2002}.

This effect is negligible at low redshifts; however, at high redshifts, it could introduce an artificial acceleration component. In \cite{Plionis2011} we  discussed this effect and how to correct for it, following \cite{HolzLinder2005}, which is based on the notion of flux conservation, i.e., that the mean flux, over all different paths of a source, naturally converges to its unlensed value. Therefore, if we had a large number of standard candles densely populating all the redshift bins, the lensing effects  would be smoothed out and it would be unnecessary to correct. 

\begin{figure}
\centering\includegraphics[width=\columnwidth, trim=0 5.5cm 0 3cm, clip]{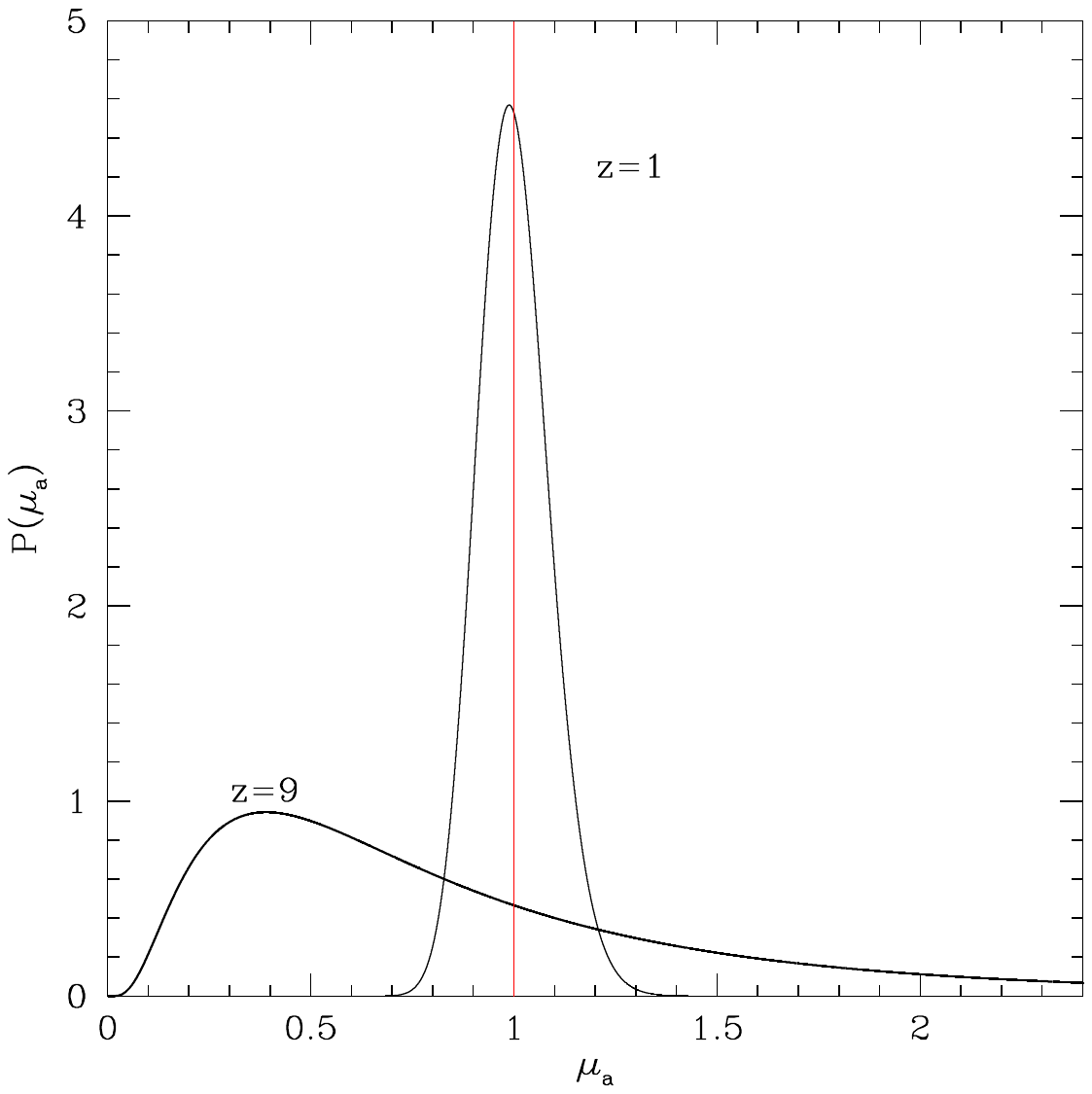}
\caption{ The magnification probability density function for a  $z=1$ and a  $z=9$ source.}
\label{fig:Len_pdf}
\end{figure}

However, we have neither a dense population of sources, nor sources that can be considered standard candles, so we must account for the lensing of each individual source. To this end we model the lensing magnification as a latent variable and marginalise over it in the likelihood. The magnification probability distribution function, used in our likelihood analyses, is a log-normal, as discussed previously, and it becomes skewed and broadened with increasing redshift \citep{Plionis2011,Marra2013L,Takahashi2011}. As an illustration, we present in Figure \ref{fig:Len_pdf} the $P(\mu_{a})$ at two redshifts ($z=1$ and $z=9$).

For each source at redshift \(z_i\), we denote the lensing magnification by \(\mathcal{M}_i\). We assume that \(\mathcal{M}_i\) follows a log-normal distribution, equivalently,
\begin{equation}
    t_i \equiv \ln \mathcal{M}_i \sim \mathcal{N}(U_i,S_i^2).
\end{equation}
The width of this prior is set by an effective magnification scatter, \(\sigma_{\rm eff}(z)\), such that
\begin{equation}
    \sigma_{\rm eff}(z)= 0.088\,z,
\end{equation}
(according to \cite{HolzLinder2005}) 
and therefore
\begin{equation}
    S_i=\sqrt{\ln\!\left[1+\sigma_{\rm eff}^2(z_i)\right]},
    \qquad
    U_i=-\frac{1}{2}S_i^2.
\end{equation}
This choice ensures flux conservation, \(\langle \mathcal{M}_i\rangle=1\). The adopted log-normal description is a convenient approximation to the non-Gaussian magnification PDF seen in ray-tracing simulations, and captures the dominant weak-lensing effects relevant for current standard-candle analyses \citep{HolzLinder2005,Marra2013L,Takahashi2011}.

Lensing modifies the observed flux according to
\begin{equation}
    f_i^{\rm obs}=\mathcal{M}_i\,f_i^{\rm true},
\end{equation}
which induces a shift in the observed distance modulus,
\begin{equation}
    \Delta\mu_{{\rm lens},i}
    =-2.5\log_{10}\mathcal{M}_i
    =-b\,t_i,
    \qquad
    b\equiv \frac{2.5}{\ln 10}.
\end{equation}

Using the notation of Eq.~\ref{eq:muo}, the residual entering the likelihood for object \(i\) is
\begin{equation}
    \Delta_i \equiv \mu_{o,i}-\mu_{\theta,i},
\end{equation}
where \(\mu_{o,i}\) is the distance modulus inferred from the \(L\!-\!\sigma\) relation and \(\mu_{\theta,i}\equiv \mu_\theta(z_i\mid\theta)\) is the cosmological prediction. For a fixed latent magnification, the corresponding unlensed residual is
\begin{equation}
    \Delta_i^{\,\rm unlensed}(t_i)=\Delta_i+b\,t_i.
\end{equation}
The total Gaussian core width is
\begin{equation}
    \varepsilon_i^2
    =
    \varepsilon_{\mu_o,i,\mathrm{stat}}^2
    +
    \varepsilon_{\mu_\theta,i,\mathrm{stat}}^2
    +
    \varepsilon_{\mathrm{sys},i}^2,
\end{equation}
in direct correspondence with Eq.~\ref{eq:epsilon}.

For each object within the redshift interval where the lensing correction is applied, the per-object likelihood is marginalised over the log-normal magnification prior:
\begin{equation}
\begin{aligned}
\mathcal{L}_i
&=
\int_0^\infty d\mathcal{M}_i\,
\frac{1}{\mathcal{M}_i\,S_i\sqrt{2\pi}}
\exp\!\left[
-\frac{(\ln\mathcal{M}_i-U_i)^2}{2S_i^2}
\right] \\
&\qquad \times
\exp\!\left[
-\frac{1}{2}
\frac{\left(\Delta_i+b\ln\mathcal{M}_i\right)^2}{\varepsilon_i^2}
\right].
\end{aligned}
\label{eq:lens_like_lognormal}
\end{equation}
Equivalently, in terms of \(t_i=\ln\mathcal{M}_i\),
\begin{equation}
    \mathcal{L}_i
    =
    \int_{-\infty}^{+\infty}
    \frac{dt_i}{\sqrt{2\pi}\,S_i}
    \exp\!\left[-\frac{(t_i-U_i)^2}{2S_i^2}\right]
    \exp\!\left[
    -\frac{1}{2}
    \frac{\left(\Delta_i+b\,t_i\right)^2}{\varepsilon_i^2}
    \right].
    \label{eq:lens_like_t}
\end{equation}
For objects outside the adopted lensing mask, the likelihood reduces to the standard Gaussian form,
\begin{equation}
    \ln \mathcal{L}_i
    =
    -\frac{1}{2}\frac{\Delta_i^2}{\varepsilon_i^2}
    + \mathrm{const}.
\end{equation}

In the code, marginalization in Eq.~\ref{eq:lens_like_t} is evaluated numerically using Gauss--Hermite quadrature. Although the present log-normal model also admits an analytic treatment after the change of variable \(t_i=\ln\mathcal{M}_i\), the quadrature approach is convenient in practice and straightforward to generalise to more flexible lensing PDFs \citep{Marra2013L,Takahashi2011}.

For a quadrature of order \(N_{\rm GH}=20\), with Hermite nodes and weights \(\{x_k,w_k\}_{k=1}^{N_{\rm GH}}\), we write
\begin{equation}
    t_{ik}=U_i+\sqrt{2}\,S_i\,x_k,
\end{equation}
so that
\begin{equation}
    \mathcal{L}_i
    \propto
    \frac{1}{\sqrt{\pi}}
    \sum_{k=1}^{N_{\rm GH}}
    w_k
    \exp\!\left[
    -\frac{1}{2}
    \left(
    \frac{\Delta_i+b\,t_{ik}}{\varepsilon_i}
    \right)^2
    \right].
    \label{eq:lens_GH}
\end{equation}
Accordingly,
\begin{equation}
    \ln \mathcal{L}_i
    \approx
    \ln\!\left[
    \frac{1}{\sqrt{\pi}}
    \sum_{k=1}^{N_{\rm GH}}
    w_k
    \exp\!\left(
    -\frac{1}{2}
    \left(
    \frac{\Delta_i+b\,t_{ik}}{\varepsilon_i}
    \right)^2
    \right)
    \right]
    +\mathrm{const},
\end{equation}
which is evaluated in practice using a numerically stable log-sum-exp form.

The total lensing-corrected likelihood contribution is then
\begin{equation}
    \chi^2_{\rm like}=-2\sum_i \ln \mathcal{L}_i.
\end{equation}

\begin{table*}
        \caption{Marginalised best-fit parameter values and associated $1\sigma$ uncertainties for the HIIGs and anchor samples. Values enclosed in parentheses indicate parameters that were held constant during the analysis.}
\label{tab:const}
\resizebox{\textwidth}{!}{
\begin{tabular}{@{}lcccccccccccc@{}}
\toprule
	Data Set & Model & $\alpha$ & $\beta$ & $h$ & $\Omega_m$ & $\Omega_\mathrm{DE}$ & $\Omega_k$ & $w_0$ & $w_a$  & BIC & AIC & N\\
\midrule
	HIIG & $\Lambda\mathrm{CDM}$ &--- & ($5.022\pm 0.058$) & --- & $0.273^{+0.037}_{-0.045}$ & $0.727^{+0.045}_{-0.037}$ & --- & (-1.0) & (0.0) & 544.24 & 540.90 & 207\\
    HIIG & $o\Lambda\mathrm{CDM}$ &--- & ($5.022\pm 0.058$) & --- & $0.265^{+0.042}_{-0.048}$ & $0.59^{+0.27}_{-0.16}$ & $0.14^{+0.16}_{-0.28}$ & (-1.0) & (0.0) & 549.55 & 542.88 & 207\\
	HIIG & $w\mathrm{CDM}$ &--- & ($5.022\pm 0.058$) & --- & $0.266^{+0.10}_{-0.061}$ & $0.734^{+0.061}_{-0.10}$ & --- & $-1.11^{+0.54}_{-0.29} $ & (0.0) & 549.53 & 542.86 & 207\\
    HIIG & $ow\mathrm{CDM}$ &--- & ($5.022\pm 0.058$) & --- & $0.262^{+0.079}_{-0.060}$ & $0.54^{+0.27}_{-0.22}$ & $0.19^{+0.18}_{-0.29}$ & $-1.18\pm 0.44 $ & (0.0) & 554.86 & 544.86 & 207\\
\midrule
	HIIG & $\Lambda\mathrm{CDM}$ &($33.268\pm 0.083$) & ($5.022\pm 0.058$) & $0.716\pm 0.018$ & $0.262^{+0.038}_{-0.047}$ & $0.738^{+0.047}_{-0.038}$ & --- & (-1.0) & (0.0) & 479.20 & 472.54 & 207\\
    HIIG & $o\Lambda\mathrm{CDM}$ &($33.268\pm 0.083$) & ($5.022\pm 0.058$) & $0.712\pm 0.019$ & $0.257^{+0.042}_{-0.048}$ & $0.62^{+0.29}_{-0.14}$ & $0.12^{+0.15}_{-0.29}$ &(-1.0) & (0.0) & 484.54 & 474.54 & 207\\
	HIIG & $w\mathrm{CDM}$ &($33.268\pm 0.083$) & ($5.022\pm 0.058$) & $0.718\pm 0.020$ & $0.265^{+0.094}_{-0.057}$ & $0.735^{+0.057}_{-0.094}$ & --- & $-1.17^{+0.46}_{-0.39}$ & (0.0) & 484.53 & 474.54 & 207\\
    HIIG & $ow\mathrm{CDM}$ &($33.268\pm 0.083$) & ($5.022\pm 0.058$) & $0.713\pm 0.020$ & $0.260^{+0.075}_{-0.059}$ & $0.56^{+0.27}_{-0.20}$ & $0.18^{+0.17}_{-0.29}$ &$-1.23\pm 0.43$ & (0.0) & 489.87 & 476.54 & 207\\
\midrule
    Anchor+HIIG & $\Lambda\mathrm{CDM}$  &$33.249 \pm 0.115$ & $5.020\pm 0.094$ & $0.724\pm 0.039$ & (0.33) & (0.67) & --- &(-1.0) & (0.0) & 831.93 & 821.45 & 243\\
	Anchor+HIIG & $\Lambda\mathrm{CDM}$ &$33.256 \pm 0.139$ & $5.013  \pm 0.114$ & $0.725 \pm 0.040$ & $0.308^{+0.043}_{-0.053}$ & $0.692^{+0.053}_{-0.043}$ & --- & (-1.0) & (0.0) & 837.44 & 823.47 & 243 \\ 
    Anchor+HIIG & $o\Lambda\mathrm{CDM}$ &$33.280 \pm 0.140$ & $4.993 \pm 0.114$ & $0.726\pm 0.040$ & $0.296^{+0.048}_{-0.056}$ & $0.48^{+0.26}_{-0.22}$ & $0.23^{+0.22}_{-0.28}$ &(-1.0) & (0.0) & 842.58 & 825.12 & 243 \\ 
	Anchor+HIIG & $w\mathrm{CDM}$ &$33.267 \pm 0.140$ & $5.004 \pm 0.114$ & $0.727 \pm 0.040$ & $0.266^{+0.12}_{-0.079}$ & $0.734^{+0.079}_{-0.120}$ & --- &$-0.96^{+0.53}_{-0.21}$ & (0.0) & 842.47 & 825.00 & 243\\ 
    Anchor+HIIG & $ow\mathrm{CDM}$ &$33.290 \pm 0.138$ & $4.985 \pm 0.113$ & $0.727\pm 0.039$ & $0.273^{+0.091}_{-0.060}$ & $0.45^{+0.22}_{-0.31}$ & $0.27^{+0.24}_{-0.27}$ & $-1.07^{+0.58}_{-0.45}$ & (0.0) & 848.03 & 827.07 & 243\\ 
    Anchor+HIIG & CPL &$33.269\pm 0.139$ & $5.003\pm 0.114$ & $0.726\pm 0.039$ & $0.285^{+0.11}_{-0.066}$ & $0.715^{+0.066}_{-0.110}$ & --- &$-0.92^{+0.57}_{-0.34}$ & $-0.48^{+0.60}_{-1.5}$ & 847.09 & 826.13 & 243\\ 
    Anchor+HIIG & $o$CPL &$33.291 \pm 0.138$ & $4.985 \pm 0.113$ & $0.727\pm 0.039$ & $0.275^{+0.094}_{-0.059}$ & $0.44^{+0.21}_{-0.30}$ & $0.28\pm 0.24$ & $-1.04^{+0.63}_{-0.47}$ & $-0.25^{+0.88}_{-1.7}$ & 852.64 & 828.19 & 243\\ 
\hline
\end{tabular}}
\end{table*}


\begin{figure*}
\begin{center}
\includegraphics[width=1.6\columnwidth]{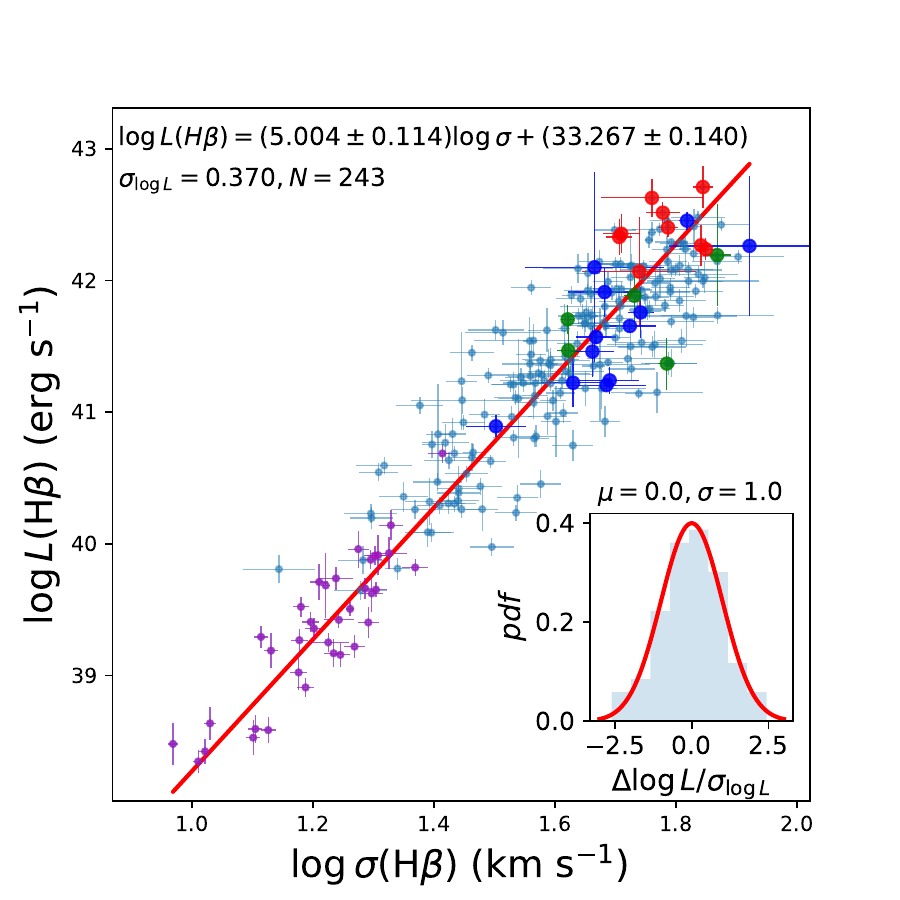}
\end{center}    
\caption{The \lsigG\ relation for the HIIGs and anchor samples. The anchor sample (36 objects; \citet{Fernandez2018}) is shown in magenta, the local-to-intermediate-$z$ HIIG sample (181 objects; \citet{2021MNRAS.505.1441G}) in light blue, 9 HIIGs from \citet{2023A&A...676A..53L} in red, and 5 HIIGs observed with JWST from \citet{2024A&A...684A..87D} in green (both the red and green subsets were presented and analysed in our previous work; \citet{Chavez2025}). The 12 HIIGs analysed in this work are shown in deep blue.
 The red line shows the best linear fit to the data, including uncertainties on both axes. The slope and intercept of the best fit and their uncertainties are shown at the top. Also shown are the standard deviation of  $\log$ L around the best fit and the total number of objects in the sample. The inset shows the pull distribution of the entire sample of GEHRs and HIIGs;  the best Gaussian fit to the distribution is shown in red.}
\label{fig:LSig}
\end{figure*}
\begin{figure}
\centering\includegraphics[width=\columnwidth]
{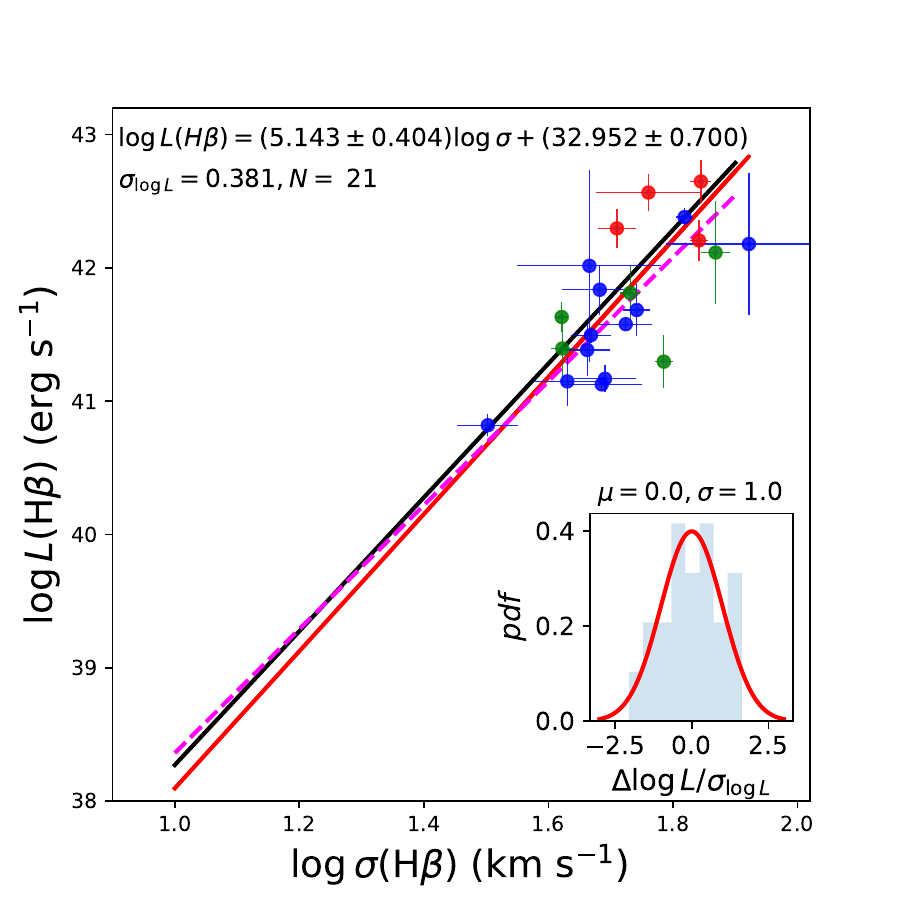}
\caption{ $L-\sigma$ relation for the 21 HIIG in our sample with z$>$3.0. The data points follow the colour coding described in the caption of Fig.~\ref{fig:LSig}. The red line shows the fit to the data. 
The slope of the fit to the high redshift sample is consistent with the slope of the anchor sample (black line) and also with that of the low redshift HIIG sample (magenta dashed line). The fit for the complete sample presented in figure \ref{fig:LSig} is not included because it coincides with the result of the fit to the 21 high redshift HIIG.}
\label{fig:Lsigma-zgt3}
\end{figure}

\begin{figure*}
\begin{center}
\includegraphics[width=1.6\columnwidth]{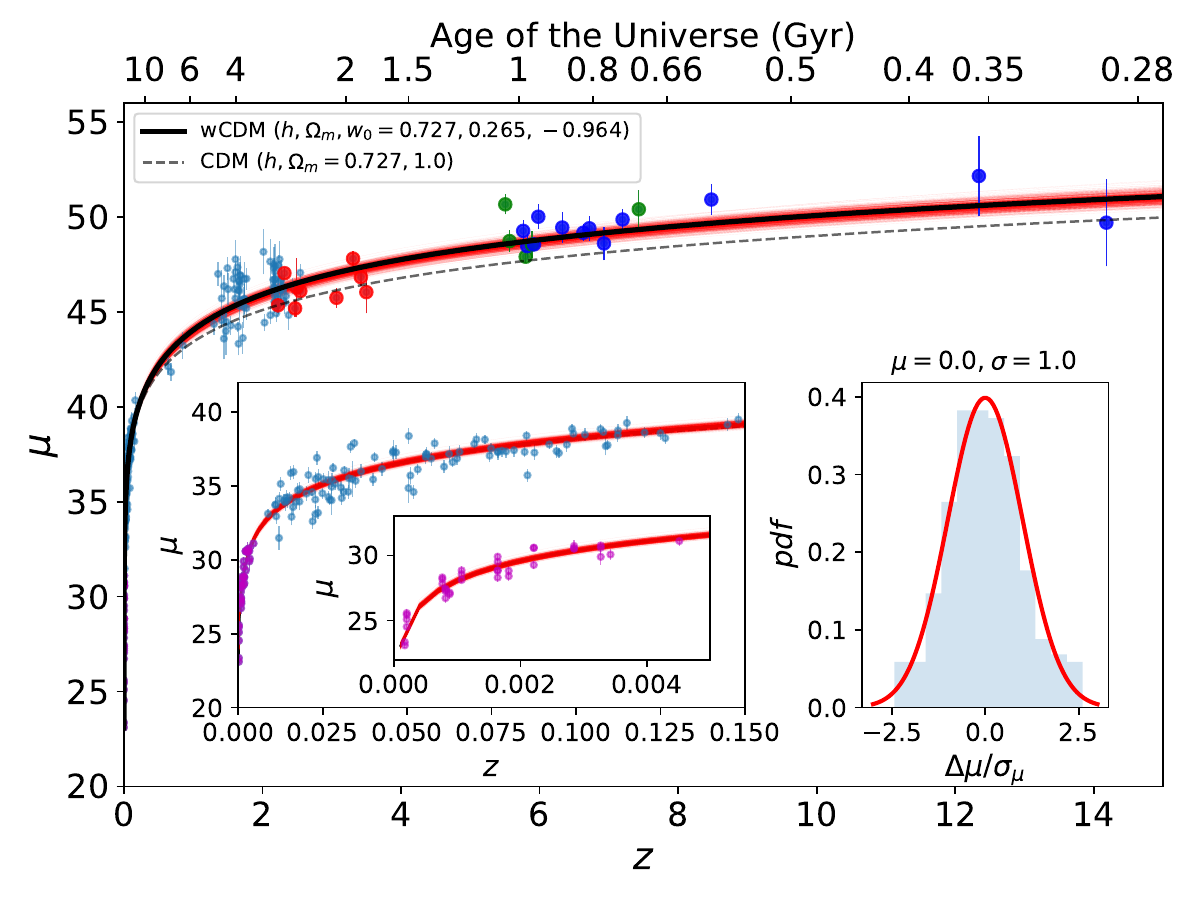}
\end{center}    
\caption{Hubble diagram for the HIIG and anchor samples, where $z$ is the redshift and $\mu$ the distance modulus. The data points follow the colour coding described in the caption of Fig.~\ref{fig:LSig}. The left inset shows a zoom on $z \leq 0.15$; the nested inset further zooms to $z \leq 0.005$, highlighting the anchor sample. The black curve indicates the best-fitting cosmological model;  the red shaded band represents the $1\sigma$ model uncertainty, and the grey dashed curve shows a flat model with no dark energy. The right inset presents the pull probability density function (pdf) for the combined GEHR$+$HIIG sample, and the red curve is the best-fitting Gaussian.}
\label{fig:hubdiag}
\end{figure*}
\begin{figure*}
\begin{center}
\includegraphics[width=1.6\columnwidth]{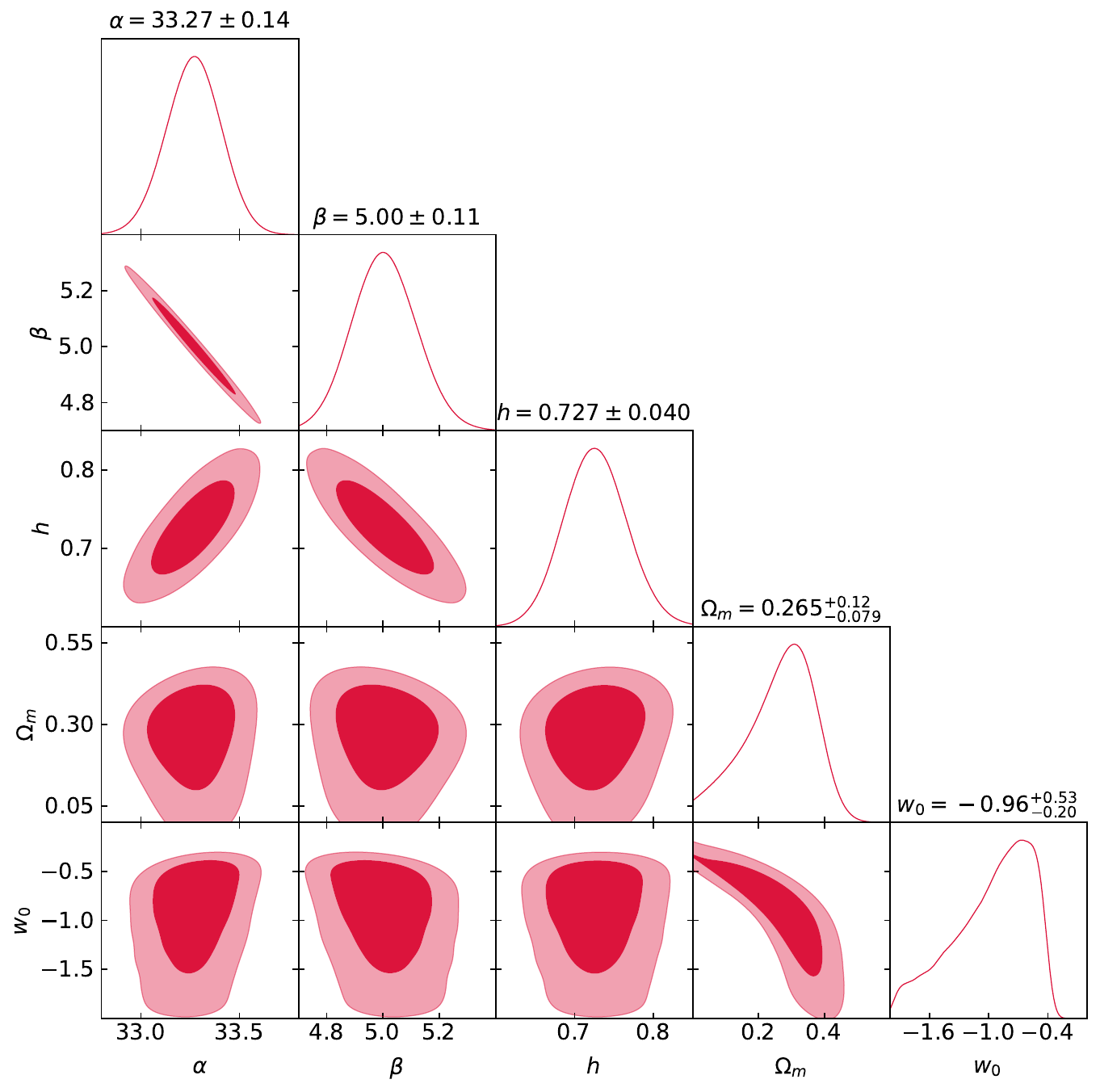}
\end{center}    
\caption{Likelihood  contours corresponding to the 1$\sigma$ and 2$\sigma$ confidence levels in the $\{\alpha, \beta, h,\Omega_m\, w_0\}$ space for the joint HIIGs and anchor samples.}
\label{fig:abhOmW0}
\end{figure*}

\section{Results}
Table~\ref{tab:const} summarises the posterior constraints  from
the 243-object compilation. Results are presented for flat and curved
versions of \(\Lambda\mathrm{CDM}\), \(w\mathrm{CDM}\), and CPL,
together with the corresponding AIC and BIC values. Parentheses indicate
parameters fixed during a given run, while the remaining entries are
posterior estimates or derived quantities.

We define the information criteria as
$\mathrm{AIC}=-2\ln\mathcal{L}_{\max}+2k$ and
$\mathrm{BIC}=-2\ln\mathcal{L}_{\max}+k\ln N$, where $k$ is the
number of free parameters and $N$ is the number of data points.
Information-criterion differences are compared only between models
fitted to the same dataset.


\subsection{The extended \(\text{L}-\sigma\) relation}

In Figure \ref{fig:LSig} we present the \(L-\sigma\) relation for the full sample of HIIGs and local anchor objects. The points are grouped into the subsamples described above, allowing the local calibrators,  previously published low- and intermediate-redshift HIIGs, and  new high-redshift JWST objects to be compared within a common luminosity--velocity-dispersion plane. The solid red line shows the best-fitting linear relation, obtained by taking into account the uncertainties in both \(\log L\) and \(\log\sigma\). The corresponding slope, intercept, and their uncertainties are reported in the figure, together with the rms dispersion in \(\log L\) about the fitted relation and the total number of objects included in the fit.

The inset shows the distribution of normalised residuals, or pulls, computed relative to the best-fitting relation. The superimposed Gaussian serves as useful diagnostic of the residual distribution and indicates that the scatter is broadly consistent with the adopted error model. The persistence of a tight \(L-\sigma\) correlation across the combined GEHR and HIIG sample, including the newly added JWST sources, supports the use of this relation as a standardisable luminosity indicator over a wide redshift interval.

\subsection{Does the $L-\sigma$ relation change with look-back time?}

In their study of the \( L-\sigma \) relation for HIIGs as cosmological distance indicators, \citet{2024PhRvD.109l3527C} proposed that the relation displays a significantly flatter slope for high-$z$ HIIGs compared to their low-$z$ counterparts. This hypothesis has important implications for using HIIGs to determine cosmological distances. However, the flattening of the $L-\sigma$ slope at high-$z$ reported by \cite{2024PhRvD.109l3527C} is directly caused by the limited luminosity range of their adopted data set, which results from the observational constraints of the higher redshift sample of \cite{GonzalezMoran2019}  employed in \cite{2024PhRvD.109l3527C}. The narrow dynamical range in luminosity, combined with  scatter in data points comparable to the luminosity range, leads to a flatter fitted slope. This idea is supported by analyses showing that the $L-\sigma$ slopes for high-$z$ and suitably matched low-$z$ samples are statistically indistinguishable when compared within similar luminosity ranges.

The extended sample presented in this paper allows a better comparison of the slopes of the different subsamples used: the anchor sample of \cite{Fernandez2018} at  $z < 0.0045$, the low-redshift subsample of \cite{Chavez2014} at $ 0.01 < z < 0.16$ and the highest redshift subsample at $z > 3$.\\

The corresponding fits are:
\begin{equation}\label{eta1}
\log L(H\beta)= (5.02\pm0.21)\log\sigma + (33.25\pm 0.25)
\end{equation}
for the anchor sample \citep{Fernandez2018},
\begin{equation}\label{eta2}
\log L(H\beta)=   (4.65\pm0.14)\log\sigma + (33.71\pm 0.21)
\end{equation}
for the low redshift sample \citep{Chavez2014} and,
\begin{equation}\label{eta3}
\log L(H\beta)=   (5.14\pm0.40)\log\sigma + (32.95\pm 0.70)
\end{equation}
for the high redshift sample in this paper, $z > 3.0$.\\

These results are illustrated in figure \ref{fig:Lsigma-zgt3} overlapping  the 21 high redshift objects. Clearly there is  no sign of evolution in the slope of the  \( L-\sigma \) relation.

It is also relevant to compare these results with the fit for the complete sample presented in figure \ref{fig:LSig}:
\begin{equation}\label{eta4}
\log L(H\beta)=   (5.00\pm0.11)\log\sigma + (33.27\pm 0.14).
\end{equation}
This fit is not included in figure \ref{fig:Lsigma-zgt3} because it coincides with the result of the fit to the anchor sample.

This is a remarkable result. The fit coefficients for the slope of the \( L-\sigma \) relation in the anchor and nearest samples—which include galaxies within 20 Mpc—agree, within the uncertainties, with those of the high-redshift sample, covering $3 < z < 14.2$.

\subsection{The Hubble diagram and cosmological parameters}

We focus on constraining a generalised parameter space, denoted as $\theta = \{\alpha, \beta, h, \Omega_m, \Omega_{\mathrm{DE}}, w_0, w_a \}$.  $\theta_n = \{\alpha, \beta \}$ represents nuisance parameters, specifically characterising the \(\text{L}-\sigma\) relation for GEHRs and HIIGs, where $\alpha$ is the intercept and $\beta$ the slope. The remainder of the parameter space concerns distinct cosmological models. For the flat $\Lambda$CDM model, the parameters are defined as $\theta_c = \{h, \Omega_m, -1, 0 \}$, indicating that we constrain the reduced Hubble constant, \(h\), and the total matter density parameter, $\Omega_m$, while keeping the first two DE EoS parameters fixed at $w_0 = -1$ and $w_a = 0$. This setting aligns with a cosmological constant ($\Lambda$). For non flat $\Lambda$CDM, or  $o\Lambda$CDM, we have the set of free parameters $\theta_c = \{h, \Omega_m, \Omega_{\mathrm{DE}}, -1, 0 \}$, where $\Omega_{\mathrm{DE}}$ is the dark energy density parameter. In an analogous way, whenever we consider a non flat case we use the prefix ``$o$" before the name of the model. Extending the constraints to include the value of $w_0$ allows for an evolving DE EoS, characteristic of models akin to quintessence \citep{PhysRevD.37.3406, Wetterich1988}. Finally, incorporating a constraint on $w_a$ aligns with the Chevallier-Polarski-Linder (CPL) model, as described in the seminal papers \citep{Chevallier2001, Linder2003, Peebles2003}.

In the first four rows of Table \ref{tab:const} we use our \( h \) free approach, so $\alpha$ and $h$ are not included in the analysis. We fix the value of $\beta$ as discussed above, hence we do not employ the anchor sample. In the following four rows, we include constraints on $h$ but fix the nuisance parameters, $\alpha$ and $\beta$,  again not employing the anchor sample. In the last seven rows of the table, we show the results including the anchor sample, so that we are constraining  nuisance and cosmological parameters simultaneously. The fact that in all cases the  nuisance parameters, $\alpha$ and $\beta$, are consistent at the  $1\sigma$ level shows the stability of the analysis and the constraints on the \(\text{L}-\sigma\) relation.

Figure \ref{fig:hubdiag} shows the Hubble diagram for the anchor and HIIG samples, with redshift \(z\) plotted against distance modulus \(\mu\). The anchor sample (36 objects; \citet{Fernandez2018}) is shown in magenta, the local-to-intermediate-\(z\) HIIG sample (181 objects; \citet{2021MNRAS.505.1441G}) in light blue, 9 HIIGs from \citet{2023A&A...676A..53L} in red, and 5 HIIGs observed with JWST from \citet{2024A&A...684A..87D} in green. The latter two sets were already presented and studied in our earlier work \citep{Chavez2025}. The 12 HIIGs analysed in this work are shown in deep blue. The left inset zooms into the \(z \leq 0.15\) region, with a further inset for \(z \leq 0.005\) highlighting the anchor sample. The black curve indicates the best-fitting cosmological model, the red shaded band its \(1\sigma\) uncertainty, and the grey dashed curve, a flat model without dark energy. The right inset shows the pull probability density function for the combined GEHR+HIIG sample, with the best-fitting Gaussian shown in red.

Figure \ref{fig:abhOmW0} shows the likelihood contours for 1$\sigma$ and 2$\sigma$ derived from a global fit to our HIIG and anchor samples. This fit includes all free parameters, both nuisance and cosmological, within a model featuring an evolving DE EoS parameter. The resulting parameter space constraints, as deduced from this figure and   Table \ref{tab:const}, are highly consistent with other recent determinations in the field \citep{Scolnic2018, 2022ApJ...938..110B}. A comparative analysis of the Figure of Merit ($FoM$), as defined in equation \ref{eqFOM}, with that reported in \citet{2021MNRAS.505.1441G}, for the common parameter space
$(\alpha,\beta,h,\Omega_m,w_0)$, reveals a notable improvement of approximately 19.3\% in our present results.

\section{Discussion and conclusions}
The inclusion of galaxies beyond \(z=8\) extends the empirical test of
the \(L-\sigma\) relation to a substantially earlier phase of cosmic
history. Within the present uncertainties, these objects follow the
same relation as the local anchors and lower-redshift HIIGs. We
therefore find no detectable change in the photo-kinematic relation over
the interval examined, although larger high-redshift samples will be
required to test subtler evolutionary effects.

The resulting cosmological constraints remain consistent with
concordance cosmology. Their principal contribution is the use of a single
independent distance indicator over an exceptionally broad redshift
baseline rather than greater precision compared to current SNIa measurements.

For the joint anchor+HIIG sample, the flat $\Lambda$CDM model
provides the lowest AIC and BIC values. The corresponding non-flat
models yield larger information-criterion values, while their curvature
posteriors remain consistent with $\Omega_k=0$. Therefore, the present
data do not provide sufficient evidence to justify the additional
curvature parameter.



The constraints on cosmological parameters derived from our dataset, as shown in Table \ref{tab:const} and Figure \ref{fig:abhOmW0}, particularly our constraints on the space $\lbrace h, \Omega_m, w_0\rbrace = \lbrace 0.727\pm0.040, 0.266^{+0.12}_{-0.079}, -0.96^{+0.53}_{-0.21}\rbrace $ (stat) from GHIIR and HIIG alone, are closely aligned with the latest results from the Pantheon+ analysis of 1550 SNIa $\lbrace h, \Omega_m, w_0\rbrace = \lbrace 0.735\pm0.011, 0.334\pm0.018, -0.90\pm0.14\rbrace $ \citep{2022ApJ...938..110B}. This concordance underscores the robustness and relevance of our findings in the broader context of contemporary cosmological research.

The immediate priority is to increase the number of objects with
high-resolution rest-frame optical spectroscopy and to improve the
characterisation of secondary parameters such as size, age, abundance,
and excitation. These observations will determine whether the present
scatter is predominantly observational or reflects additional physical
dependencies of the \(L-\sigma\) relation.


The JWST data enabled the inclusion of the most distant HIIGs observed to date in our dataset, which not only enhances statistical analysis but also significantly extends the accessible redshift range, enabling further tests of the consistency of the $\Lambda$CDM model and the examination of potential dark energy evolution across the broadest cosmic span. Moreover, this approach helps overcome biases in existing datasets that arise from their limited redshift coverage. The possibility of using a single independent distance indicator over this vast period of cosmic history, combined with an expanded dataset, is essential to reduce statistical uncertainties and tighten constraints on $\lbrace h, \Omega_m, w_0\rbrace$.

Further JWST data will enable a more detailed examination of the intrinsic properties of HIIGs. This deeper insight is essential for calibrating the \(\text{L}-\sigma\) relation and to mitigate systematic uncertainties in the derived parameters. By enhancing our understanding of the physical processes governing HIIGs, we can better interpret their \(\text{L}-\sigma\) relation, which is crucial for the independent determination of cosmological parameters.

Our analysis offers novel insights into the evolution of the Universe. By leveraging this new data, we significantly enhance the existing narrative of cosmic history. Our results deepen our understanding of the early Universe conditions and their role in the formation and evolution of galaxies, setting the stage for further exploration of the photokinematic properties of these massive regions of star formation.

\section*{Data availability}
The datasets supporting the conclusions of this article, including the data used for generating the figures, are available in the references given in the Data Sets section and from the corresponding author upon reasonable request.

\section*{Code availability}
AstroPy \cite{astropy:2022}, Multinest \cite{Feroz2009, 2019OJAp....2E..10F} and PyMultinest \cite{2014A&A...564A.125B}, are all publicly available, while the code used for the data analysis and figure generation for this article is publicly available via GitHub at \url{https://github.com/blackdragonae/hiigs}

\section*{Acknowledgments}
The authors are grateful to the anonymous referee whose comments improved the clarity of this paper. ET, RT and SZ acknowledge support from grant PID2022-136598NB-C33 funded by MCIN/AEI/10.13039/501100011033 and by “ERDF A Way of Making Europe”. SZ acknowledges support from the European Union (ERC, WINGS, 101040227). ALG acknowledges support from the SECIHTI grant from the program “EPM 2025”. FDE and XJ acknowledge support by the Science and Technology Facilities Council (STFC), by the ERC through Advanced Grant 695671 ``QUENCH'', and by the UKRI Frontier Research grant RISEandFALL. RA acknowledges support of Grant PID2023-147386NBI00 funded by MICIU/AEI/10.13039/501100011033 and by ERDF/EU, from grant PID2022-136598NBC32 “Estallidos8” and the Severo Ochoa award to the IAA-CSIC CEX2021-001131S, funded by MICIU/AEI/10. 13039/501100011033.


\bibliography{bib/bib2022}







\bsp	
\label{lastpage}
\end{document}